\documentclass{article}
\usepackage{arxiv}
\usepackage[utf8]{inputenc} 
\usepackage[T1]{fontenc}    
\usepackage{hyperref}       
\usepackage{url}            
\usepackage{booktabs}       
\usepackage{amsfonts}       
\usepackage{nicefrac}       
\usepackage{microtype}      
\usepackage{graphicx}
\usepackage{doi}
\usepackage{bm}
\usepackage{array}
\usepackage{eqparbox}
\usepackage{commath}
\usepackage{nicefrac}    
\usepackage{microtype} 
\usepackage[section]{placeins}
\usepackage{hhline}
\usepackage{cite}
\usepackage{subfigure}
\usepackage{multirow}
\usepackage{array}
\usepackage{mdwmath}
\usepackage{mdwtab}
\usepackage{amssymb,amsmath,amsthm,mathtools}
\usepackage{enumerate}
\usepackage{setspace}
\usepackage{comment}
\usepackage{cleveref}
\usepackage{float}
\usepackage{makeidx}
\usepackage{algorithm2e}
\usepackage{xcolor}

\title{On the Use of Safety Critical Control for Cyber-Physical Security in {the Smart Grid}}

\date{} 					

\author{{Amr S.~Mohamed} \hspace{7mm} {Deepa~Kundur}\\
	Department of Electrical Engineering\\
	University of Toronto\\
	Toronto, ON M5S 3G4, Canada \\
	\texttt{amr.mohamed@mail.utoronto.ca} \\
 \texttt{dkundur@ece.utoronto.ca} \\
        \And
	Mohsen Khalaf \\
         Department of Electrical Engineering\\
	University of Toronto\\
 Toronto, ON M5S 3G4, Canada \\
	and Assiut University\\
	Assiut, Egypt\\
	\texttt{m.khalaf@utoronto.ca}\\
        \texttt{mohsen.b.khalaf@aun.edu.eg} \\
}

\renewcommand{\shorttitle}{On the Use of Safety Critical Control for Cyber-Physical Security in {the Smart Grid}}

\begin{document}
\maketitle

\begin{abstract}
The tight coupling between communication and control in cyber-physical systems is necessary to enable the complex regulation required to operate these systems. Unfortunately, cyberattackers can exploit network vulnerabilities to compromise communication and force unsafe decision-making and dynamics. If a cyberattack is not detected and isolated in a timely manner, the control process must balance adhering to the received measurement signals to maintain system operation and ensuring that temporary compromise of the signals does not force unsafe dynamics. For this purpose, we present and employ a safety critical controller based on control barrier functions to mitigate attacks against load frequency control in smart power grids. {We validate the paper's findings using simulation on a high-fidelity testbed}. 
\end{abstract}

\keywords{
cyber-physical security, barrier control function, safety-critical control, cyberattack, attack mitigation}

\section{Introduction} \label{sec:1}

Cyber-physical systems involve a tight coupling between the cyber layer comprised of computation and communication, and the physical layer of control, actuation and physical dynamics. 
By exploiting cyber vulnerabilities, malicious attackers can mislead control {-- inducing} unsafe physical dynamics, equipment damage and/or system destabilization and failure. 
For example, the Stuxnet \cite{langner2011stuxnet} and Aurora attack \cite{greenberg2019sandworm} demonstrate how compromising control can lead to equipment damage. 

In power systems, a key requirement of control is to maintain system security, which refers to sustaining reliable power supply and equipment safety in the face of unexpected and sudden disruptions in real time, such as faults, unanticipated loss of generation or network components, or rapid changes in demand \cite{iea2022strengthening}. 
In maintaining system security, protection devices are deployed to protect equipment from any severe damage due to fault current or abnormal unsafe operating conditions.
By misleading control, research has shown that cyberattacks can force unsafe dynamic behavior that falsely triggers protection devices leading to loss of generation, system instability and cascading failure. 
The authors in \cite{wu2017resonance} presented resonance attacks, in which the attacker corrupts power demand signals to induce unstable frequency oscillations that falsely trigger rate-of-change of frequency (ROCOF) relays deployed to protect generation equipment from operating in unsafe frequency. 
{Similarly, to trip ROCOF protection, authors in \cite{mohamed2021false} compromised microgrid synchronization, and authors in \cite{brown2017risk} compromised the control of loads providing emulated inertia services.
Targeting the false operation of under-frequency load shedding (UFLS) relays, the authors in \cite{khalaf2018false} manipulated the frequency measurements used in wide-area UFLS schemes. 
The authors in \cite{jafari2021false} discussed how ROCOF and load shedding relays can be triggered by manipulating distributed energy resources' measurements that are used by the system operator to estimate power generation capacity, 
Relays can also be triggered by misleading Automatic Generation Control, which the authors in \cite{khalaf2018joint} did by manipulating frequency and power flow measurements. 
}

In these attacks, we observe a control conflict: 
control relies on the communicated control signals to maintain the system security of the complex power system; 
however, in the presence of an undetected cyberattack, adhering to the communicated control signals can induce unsafe physical dynamics leading to unnecessary tripping. 
Hence, we address how to design control to ensure that system stability and security objectives are met, while protecting the system from unsafe behavior. 
Recently, safety critical control (SCC) (based on control barrier functions) has seen increasing use in developing control that addresses conflicting control objectives and safety constraints. 
In this paper -- in the context of power systems, we will use safety to refer to the protection safety bounds, beyond which protection devices will trigger.

In SCC, control barrier functions are used to minimally modify existing stabilizing control to guarantee safety. It has seen increasing use in automotive and robotic applications. Ames \textit{et al.} \cite{ames2016control} applied SCC to adaptive cruise control and lane keeping in automotives. In adaptive cruise control, the stabilizing control works to maintain the vehicle's velocity at the user-specified setpoint, and the introduced safety control ensures that the vehicle maintains a safe gap to the leading vehicles. In lane keeping, the stabilizing control maintains the vehicle at the center of the lane, while the introduced safety control ensures the vehicle's acceleration does not cause passenger discomfort. 
The authors in \cite{hsu2015control} achieved stable walking of a bipedal robot by encoding physical constraints that the robot must respect in its safety controller. The use of barrier functions in power system research is still in its very early stages. The authors in \cite{vu2021barrier} applied barrier functions to train a reinforcement learning agent to take safe emergency load shedding actions. 
SCC is yet to be applied to cyber-physical security.
We see two important application of SCC in cyber-physical security: 1) mitigating cyberattacks and 2) learning attacker strategies. 

In this paper, we consider power system cyber-physical security. While the method developed in this paper is generally applicable to a myriad of control schemes in power systems, we focus our study on attacks against Load Frequency Control (LFC) for its importance to power system operation. LFC regulates the frequency of the power system, and is critical to maintaining power system security. LFC is designed to operate with minimal human intervention and relies on control signals and measurements transmitted {via open communication protocols in wide area networks}, making it prone to cyberattacks \cite{shen2019cyber}. Research \cite{wu2017resonance, mohamed2021false, brown2017risk, jafari2021false, khalaf2018joint} has shown that attacks against LFC can lead to false protection relay operation and consequentially to potential power system {instability.}
{
The proposed method differs from prior work on mitigation against LFC attacks in that it does not require prior detection of false data at the control center and estimation of correct control signals through forecasting \cite{sridhar2014model}, neural networks, faulty sensors removal \cite{tan2017modeling}, or Kalman filters \cite{khalaf2018joint}. Instead, the SCC is deployed at the generator, relying on local measurements that are highly infeasible for attackers to compromise, and only minimally adjusts incoming control signals to keep the generator from following corrupt signals that may trip it. This results in a simpler defense strategy that does not depend on accurate prior forecasts or false data detection, and strictly operates when necessary to avoid unsafe dynamics.
}

{For} the first time, we apply SCC to the cyber-physical security of power systems. We formulate a barrier function-based controller to mitigate cyberattacks targeting the false operation of frequency (including ROCOF) relays in smart grids. We validate the effectiveness of the controller in mitigating different attacks using a high-fidelity testbed on {Simulink Simscape}. We also verify that the controller does not hinder existing stabilizing control or protection in the power system. 

In Section \ref{sec2:background}, we review frequency control and protection in power systems, and safety barrier functions-based control. In Section \ref{sec3:method}, we apply safety barrier functions-based control to frequency control, and present the results in Section \ref{sec4:results}. The paper's conclusion is in Section \ref{sec5:conclusion}.

\section{Background} \label{sec2:background}

\subsection{Frequency Control and Protection}

\begin{figure}
    \centering
    \includegraphics[width=0.5\textwidth]{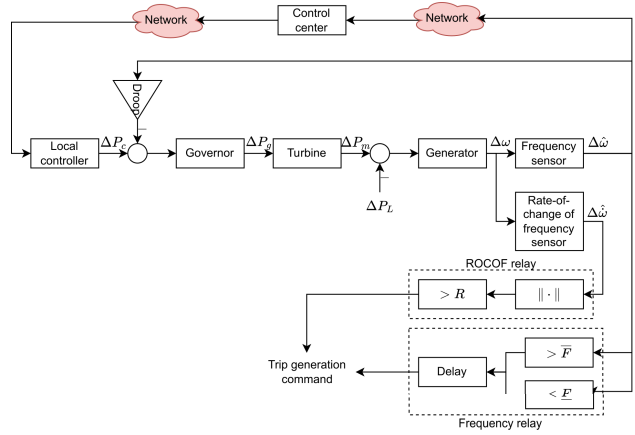}
    \caption{Microgrid load frequency control block diagram}
    \label{fig:lfc_block_dgm}
\end{figure}

We adopt the typical single-area LFC diagram \cite{kundur1994power} shown in Figure \ref{fig:lfc_block_dgm}. 
Generally, it is assumed that all generation units in a single-area can be represented by a single unit as they produce a coherent aggregate response to load change \cite{saadat1999power}. 
{The general dynamic equations for a linear single-area power system are expressed be the following state-space system.}

\begin{align} \label{eq:state-space}
    \bm{\dot{x}} &= f(\bm{x}) + g(\bm{x})(\bm{u}) = {\bm{A}} \bm{x} + {\bm{B}} \bm{u}\\ \nonumber
    \bm{x} &= \begin{bmatrix} \Delta P_g & \Delta P_m & \Delta \omega & \Delta \hat{\omega} & \hat{\dot{\omega}} \end{bmatrix}^T\\ \nonumber
    \bm{u} &= \begin{bmatrix} \Delta P_c & \Delta P_L \end{bmatrix}^T\\ \nonumber
    \bm{A} &= \begin{bmatrix} 
    -1/\tau_G & 0 & 1/(R\tau_G) & 0 & 0\\
    1/\tau_T & -1/\tau_T & 0 & 0 & 0\\
    0 & 1/M & -D/M & 0 & 0\\
    0 & 0 & 1/\tau_\omega & -1/\tau_\omega & 0\\
    0 & 1/(M\tau_\nu) & -D/(M\tau_\nu) & 0 & -1/\tau_\nu
    \end{bmatrix}\\  \nonumber
    \bm{B} &= \begin{bmatrix} 
    1/\tau_G & 0 & 0 & 0 & 0\\
    0 & 0 & -1/M & 0 & -1/(M\tau_\nu)
\end{bmatrix}^T
\end{align}
\noindent where {$\bm{x}$ and $\bm{u}$ are the state and input vectors. $\Delta$ denotes change in state/input value.} $P_{c}$ is the governor-droop control signal,
$P_{g}$ is the governor output, $P_{m}$ is the mechanical power, $\omega$ is the system frequency, $\hat{\omega}$ is the frequency measurement, and $\hat{\dot{\omega}}$ is the rate-of-change of frequency measurement. 
$\tau$ denotes the time-constants of the blocks. Subscripts $G$, $T$, $\omega$ and $\nu$ refer to the governor, turbine, frequency sensor, and rate-of-change of frequency sensor, respectively. $M$ and $D$ denote the generation's inertia and load damping constants. $R$ is the droop constant. We mathematically express the local controller as follows:
\begin{equation}
    \frac{d}{dt} \Delta P_c = k(\Delta \omega_{ref}-\Delta \hat{\omega})
\end{equation}
where $\Delta \omega_{ref}$ is the change in frequency reference set by the control center, and $k$ is some control gain.

{
The SCADA system located at the control center is responsible for LFC functions such as Automatic Generation Control \cite{kundur1994power}, microgrid synchronization, and emergency control. 
Vulnerabilities in the SCADA network or communication protocols can be exploited by cyberattackers. 
Protocols commonly used in the energy sector, such as DNP3 and IEC 61850, have security vulnerabilities due to lack of authentication, encryption, and authorization \cite{xu2017review, east2009taxonomy}, which can be exploited by cyberattackers to manipulate system measurements.}
If the attacker can force the system frequency or rate-of-change of frequency to exceed the frequency protection relay settings (characterizing the safe operating bounds), then the cyberattack can result in loss of generation, followed by potential cascading failures and/or blackout.
Figure \ref{fig:lfc_block_dgm} illustrates the ROCOF and (over- and under-) frequency relay elements which monitor the frequency to ensure that $(\hat{\omega}, \hat{\dot{\omega}})$ are within a safe set \begin{equation}
    \mathcal{S} = \{(\hat{\omega}, \hat{\dot{\omega}}): \underline{F} \leq \hat{\omega} \leq \overline{F}, \abs{\hat{\dot{\omega}}} \leq R\}
\end{equation}
where $\underline{F}, \overline{F}$ and $R$ are the settings of the under-frequency, over-frequency and ROCOF protections functions. {Recommended settings per IEEE1547 \cite{ieee1547} are listed in Table \ref{table:relaysettings}}.

\begin{table}[t!]
\centering
\caption{{IEEE1547 (Cat. III) recommended relay settings}}
\begin{tabular}[t]{lcc} 
\toprule
\textbf{Protection Function} & \textbf{Threshold (pu)} & \textbf{Clearing time}\\
\midrule
OF2 & 1.03 & 160 ms\\
UF2 & 0.942 & 160 ms\\
\midrule
ROCOF & 0.05 & \\
\bottomrule
\label{table:relaysettings}
\end{tabular}
\end{table}

\vspace{-4mm}

\subsection{Safety Barrier Functions and Control}

Safety-critical control involves minimally modifying the existing control of a system to guarantee safety. Here, safety is expressed as a safe set $\mathcal{S}$ and the system is considered safe inside the set. The safe set is defined as the superlevel set of a continuously differentiable function $h : \mathbb{R}^n \rightarrow \mathbb{R}$ yielding:
\begin{align}
    \mathcal{S} &= \{x \in \mathbb{R}^n : h(x) \geq 0 \} \nonumber\\
    \partial \mathcal{S} &= \{x \in \mathbb{R}^n : h(x) = 0 \}\\
    \text{Int}(\mathcal{S}) &= \{x \in \mathbb{R}^n : h(x) > 0 \} \nonumber
\end{align}

$h$ is a control barrier function if there exists a function $\alpha: \mathbb{R} \rightarrow \mathbb{R}$ that is strictly increasing with $\alpha(0) = 0$ such that for the control system $\dot{x} = f(x) + g(x)u$
\begin{equation}
    \label{eq:cbf}
    \dot{h} = \frac{\partial h}{\partial x} \dot{x} = \frac{\partial h}{\partial x} (f(x) + g(x)u) \geq -\alpha(h(x)) 
\end{equation}

{When the system is safe, $x$ is in the interior of $\mathcal{S}$, and $h(x)$ and $\alpha(h(x))$ are positive.} Per (\ref{eq:cbf}), $h(x)$ is allowed to decrease towards the system boundary ($h(x) = 0$) if needed. It will still remain in the safety set. Once at the boundary of $\mathcal{S}$, $\alpha(h(x)) = 0$ forcing $h$ to stay at the boundary or increase, moving away from the boundary into the interior of the safety set. Hence, the safe set $\mathcal{S}$ is rendered forward invariant by the control $u$. Proofs are presented in \cite{ames2019control}. 

Finally, the SCC can be constructed as a Quadratic program
\begin{align}
    u = \arg & \min_{u} & \frac{1}{2} \|u - u_{ref} \|_2\\
    & \text{s.t.} & \frac{\partial h}{\partial x} (f(x) + g(x)u) \geq -\alpha(h(x)) \nonumber
\end{align}

When the reference (stabilizing) control signal $u_{ref}$ results in the system respecting the safety constraint, the SCC does not change the control. Otherwise, the SCC minimally modifies the control signal to respect safety.

\subsection{Threat Model}

Cyberattacks can exploit vulnerabilities in insecure power system networks. For example, as outlined in \cite{sridhar2010data}, the attacker can gain unauthorized access to the corporate network via a social engineering attack (e.g., infected email or USB device). 
Next, the attacker can breach the power system's SCADA network through virtual private network access. 
Once breached, the attacker can corrupt communicated signals through exploiting vulnerabilities in open communication protocols that are utilized in the SCADA network for communication between the control center and {field measurement devices.} 

{We assume that the attacker has gained unauthorized network access and is able to corrupt measurements to the control center. 
The measurement corruption misleads generation control.}
As our research addresses attack mitigation, we do not assume any limitations on the attacker's knowledge of the system or attack detection strategies that would limit that attacker's ability to corrupt measurements.
The attacker's goal is to cause sufficient deviation in the system's frequency or rate of change of frequency to cause false relay operation leading to generation loss.  

\section{{SCC for LFC Cyber-physical security}} \label{sec3:method}


\begin{figure}
    \centering
    \vspace{2mm}
    \includegraphics[width=0.45\textwidth]{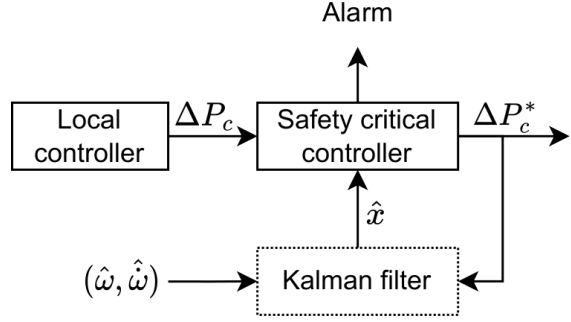}
    \caption{Safety critical controller block diagram}
    \label{fig:scc_block_dgm}
\end{figure}

We formulate the SCC by evaluating whether the current governor-droop control signal can force the system to exit the safety set $T_s$ seconds into the future. 
Consequentially, we evaluate whether the existing control can cause a ROCOF or frequency relay trigger in the next $T_s$ seconds. 
To simplify the expression for the control barrier functions, we compute $F = \min(\overline{F}-1, 1-\underline{F})$ pu {as the frequency relay bound on the deviation of frequency}. The control barrier function defining the safety set of the frequency is
\begin{equation} \label{eq:cbf_freq}
    h_\omega(\bm{x}) = F - \Delta \hat{\omega}_{t+T_s}^2(\bm{x})
\end{equation}


Similarly, the control barrier function defining the safety set of the rate-of-change of frequency is
\begin{equation} \label{eq:cbf_rocof}
    h_\nu(\bm{x}) = R - \hat{\dot{\omega}}_{t+T_s}^2(\bm{x})
\end{equation}

To compute the state $T_s$ seconds into the future, we discretize the state-space matrices (\ref{eq:state-space})
\begin{align}
    \bm{A}_d &= e^{\bm{A} T_s}\\
    \bm{B}_d &= \bm{A}^{-1} (\bm{A}_d - \bm{I}) \bm{B}
\end{align}

Hence,
\begin{equation}
    \begin{bmatrix} \Delta \hat{\omega}_{t+T_s}\\ \hat{\dot{\omega}}_{t+T_s} \end{bmatrix}(\bm{x}) = \bm{C} (\bm{A}_d \bm{x} + \bm{B}_d \bm{u} )
\end{equation}

where 
\begin{equation*}
    \bm{C} = \begin{bmatrix} \bm{C}_\omega\\ \bm{C}_\nu  \end{bmatrix} = \begin{bmatrix} 0 & 0 & 0 & 1 & 0\\ 0 & 0 & 0 & 0 & 1 \end{bmatrix}
\end{equation*}

Next, motivated by the work of the authors in \cite{ames2016control}, we compute a logarithmic barrier function based on $h$:
\begin{equation}
    \mathcal{B}(\bm{x}) = -\log \left( \frac{h(\bm{x})}{1+h(\bm{x})} \right)
\end{equation}

The SCC quadratic program is 
\begin{align}
    \Delta P_c^* = &\arg\min_{u} & \frac{1}{2} \| \Delta P_c - u \|_2 \\
    &\text{subject to} &\Dot{\mathcal{B}_\omega}(\bm{x}) - \frac{\alpha}{\mathcal{B}_\omega(\bm{x})} \leq 0 \nonumber\\
    & &\Dot{\mathcal{B}_\nu}(\bm{x}) - \frac{\alpha}{\mathcal{B}_\nu(\bm{x})} \leq 0 \nonumber
\end{align}

where 
\begin{align}
    \Dot{\mathcal{B}_\omega}(\bm{x}) &= \frac{d \mathcal{B}_\omega}{d \bm{x}} \Dot{\bm{x}} \nonumber\\
    &= - \frac{\frac{dh_\omega}{dx}}{h_\omega(\bm{x}) + h_\omega^2(\bm{x})}(\bm{A} \bm{x} + \bm{B} \bm{u} )\\
    \frac{dh_\omega}{dx} &= -2 \Delta \hat{\omega}_{t+T_s}(\bm{x}) \frac{d}{d\bm{x}} \Delta \hat{\omega}_{t+T_s} \nonumber\\ 
    &= -2 \Delta \hat{\omega}_{t+T_s}(\bm{x}) \bm{C}_\omega \bm{A}_d
\end{align}

Similarly, 
\begin{align}
    \Dot{\mathcal{B}_\nu}(\bm{x}) &= - \frac{\frac{dh_\nu}{dx}}{h_\nu(\bm{x}) + h_\nu^2(\bm{x})}(\bm{A} \bm{x} + \bm{B} \bm{u} )\\
    \frac{dh_\nu}{dx} &= -2 \hat{\dot{\omega}}_{t+T_s}(\bm{x}) \bm{C}_\nu \bm{A}_d
\end{align}

Figure \ref{fig:scc_block_dgm} illustrates the block diagram of the SCC. 
The block is implemented in the local controller, taking in the governor control signal and minimally modifying it to $\Delta P_c^*$ to guarantee safety. 
Additionally, an alarm can be produced whenever the SCC modifies the control signal. 
The system operator can inspect the communication network on alarm for breaches and take corrective actions (such as isolating the compromised communication channels).

Note that the SCC requires the state values. By observing the frequency and rate-of-change of frequency measurements, the system (\ref{eq:state-space}) is observable and a Kalman filter can be designed to estimate the state from these measurements. Due to space limitations and given the prevalence of Kalman filters, we do not cover the design of the Kalman filter here.

\section{Results and Discussion} \label{sec4:results}

\begin{figure}
    \centering
    \includegraphics[width=0.5\textwidth]{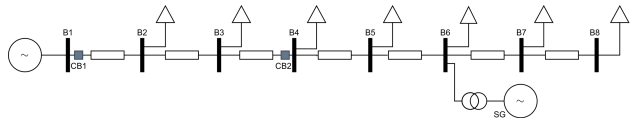}
    \caption{Microgrid testbed}
    \label{fig:mg_testbed}
\end{figure}

\begin{figure*}
    \centering
    \begin{minipage}[b]{.4\textwidth}
    
        \includegraphics[width=\textwidth]{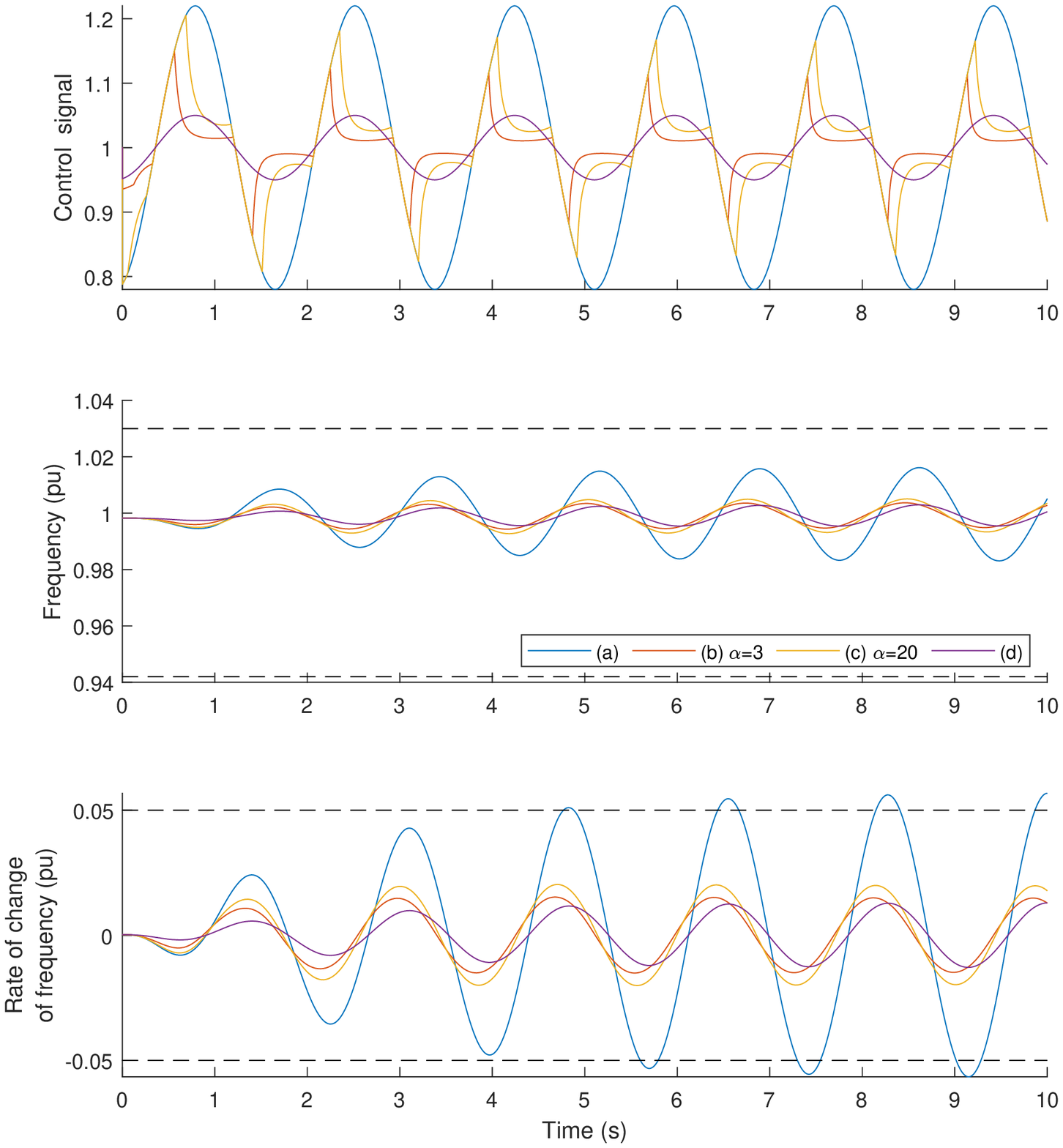}
        
        \caption{Simulation results: (a) ROCOF relay-triggering attack. SCC-based mitigation with (b) $\alpha = 3$ and (c) $\alpha = 20$. (d) Attack does not trigger ROCOF relay or activate SCC.}
        \label{fig:hypersim_rocof}
    \end{minipage}  \hfill
    \begin{minipage}[b]{.29\textwidth}
        \includegraphics[width=\textwidth]{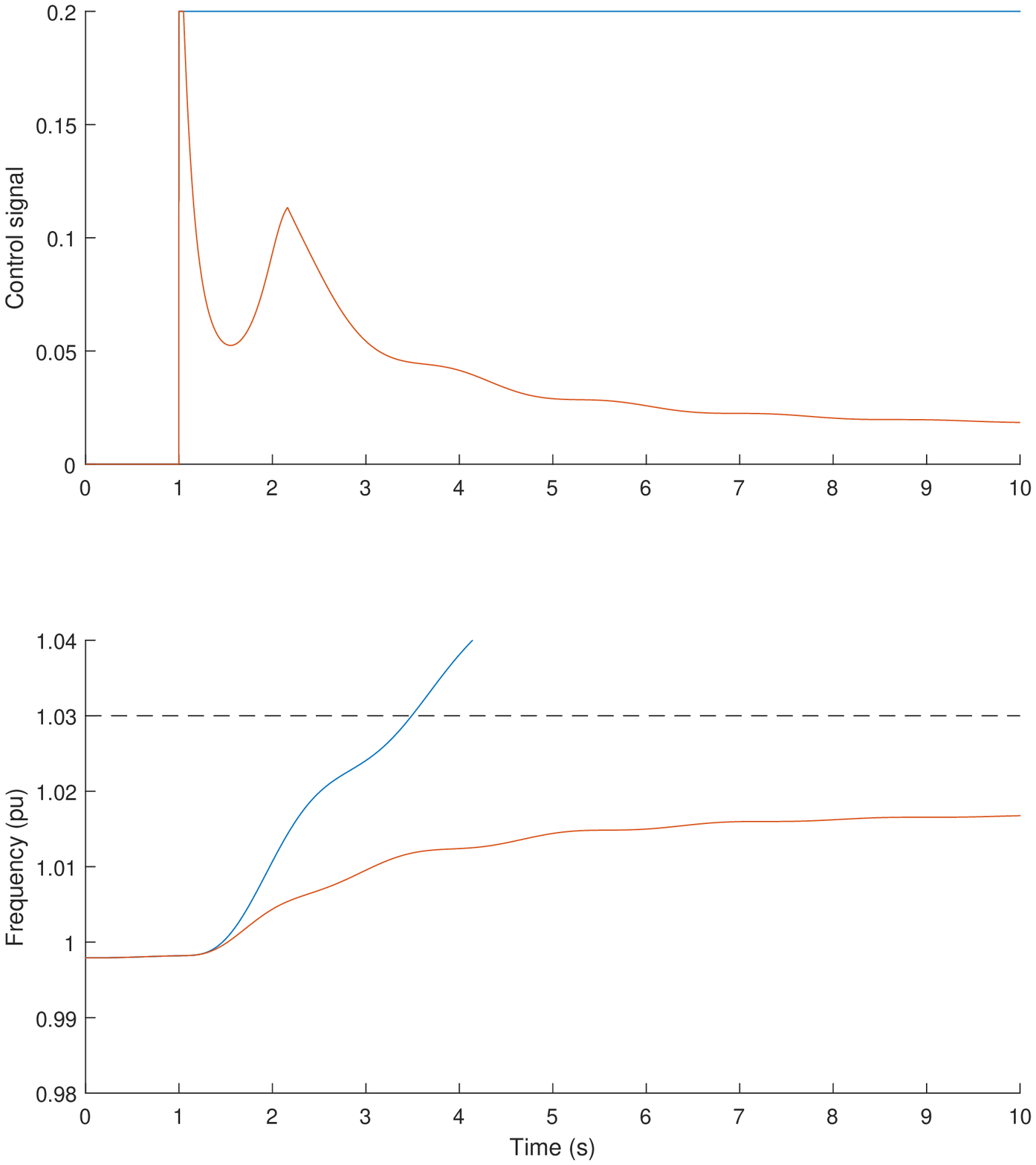}
            
        \caption{Simulation results: over-frequency relay-triggering SCC-based attack mitigation with $\alpha = 20$}\label{fig:hypersim_bias}
    \end{minipage}  \hfill
    \begin{minipage}[b]{.29\textwidth}
    
        \includegraphics[width=\textwidth]{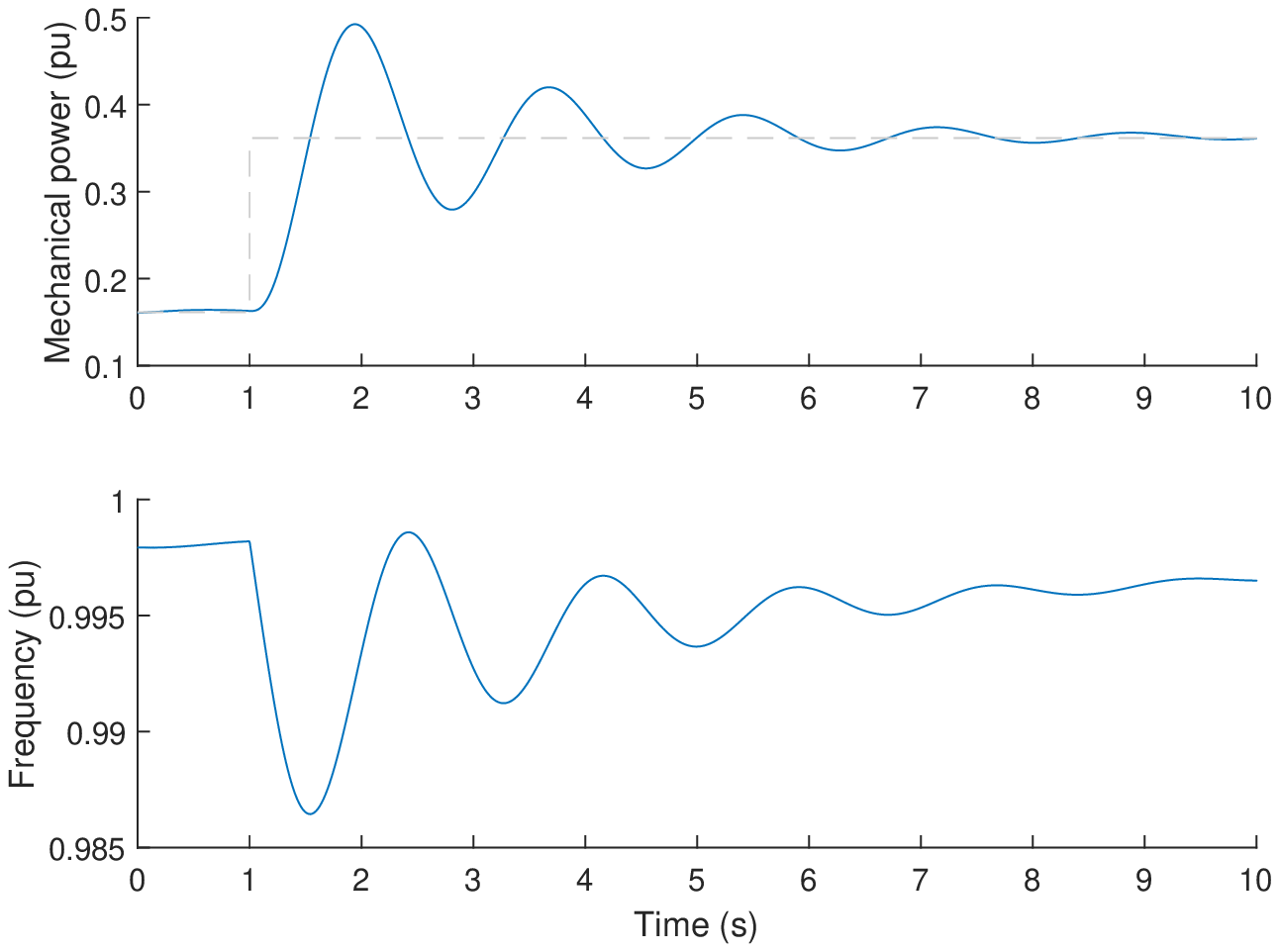}
        \caption{Simulation results: sudden addition of load and frequency stabilization}\label{fig:hypersim_load_disturbance}
        
        \includegraphics[width=\textwidth, trim={0 0 0 7cm},clip]{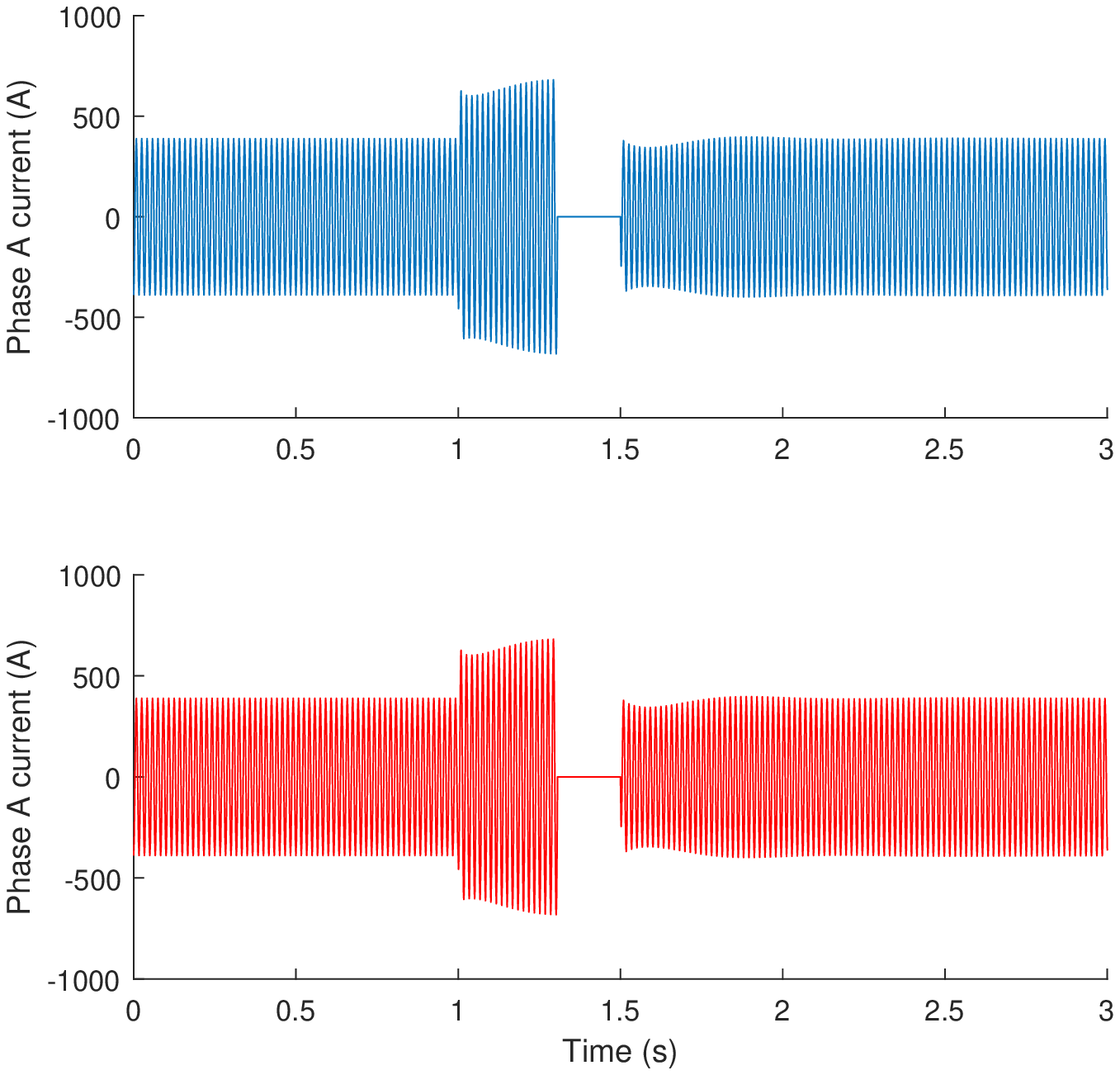}
        \caption{Simulation results: 3-phase fault at bus B5, protection and post fault clearing frequency stabilization}\label{fig:hypersim_fault}
    \end{minipage}

\end{figure*}




{In this section, we validate the method using time-domain simulations on a detailed microgrid testbed (Fig. \ref{fig:mg_testbed}) modelled with {Simulink Simscape}.
Due to their inherent low inertia, microgrids are more susceptible to cyberattacks and are ideal for testing our proposed method.
The testbed replicates a medium-voltage rural distribution system in Ontario, Canada. SG is a 2.5 MVA synchronous generator to which the SCC is applied to protect the microgrid during islanded-mode operation. 
The microgrid test system data are provided in \cite{mohamed2021false}.
The SCC applies the LFC state-space system \eqref{eq:state-space} to construct its Quadratic program.
}
Naming conventions are used such that CB and B represent circuit breakers and buses, respectively. 
We assume that the generator's frequency protection is set to $F = 0.03$ pu Hz and $R = 0.05$ pu Hz/s (per UF/OF2 and ROCOF protection functions in Table \ref{table:relaysettings}). 
{For the safety controller, we discretize the system with $T_s = 250$ milliseconds.}

\subsection{Attack Mitigation}   
\subsubsection{ROCOF relay}

{The top plot in Fig. \ref{fig:hypersim_rocof} shows (in blue (a)) a cyberattack-injected governor-droop control signal $P_{c}$ which triggers the ROCOF relay.} The triggering is evident by the (blue (a)) rate-of-change of frequency exceeding the relay setting $R$ illustrated by horizontal dashed lines in the bottom plot. 
While the ROCOF relay would trip the generation out when the rate-of-change of frequency first intersects the dashed lines, we continue to show the effect of the attack for demonstrating the SCC's continuity in maintaining safety. 
Curves (b) and (c) in Fig. \ref{fig:hypersim_rocof} show the effect of the SCC when $\alpha = 3$ and $20$, respectively. The rate-of-change of frequency is maintained within the safety bounds -- preventing the attack's success. 
Decreasing $\alpha$ results in stricter safety control: the smaller value of $\alpha$ maintains a larger safety margin.


When the attack is weaker than to trigger protection and risk system safety as in curves (d) in Fig. \ref{fig:hypersim_rocof}, SCC does not modify the control signal.

Note that while the compromised control signal is sinusoidal in our simulations, the results are general to other oscillatory signals (e.g., square, rectangular, etc.)

\subsubsection{Frequency relay}

Figure \ref{fig:hypersim_bias} shows (in blue) a cyberattack which triggers the over-frequency relay. 
The triggering is evident by the frequency exceeding the OF2 relay setting illustrated by the horizontal dashed line.
The SCC prevents the attack's success by maintaining the frequency within the safety bounds.

\subsection{Learning Attacker Strategies}

Reliable attack detection measures can prevent the success of the discussed attack by isolating the compromised communication channels following detection. 
However, if the cyberattack manages to circumvent these measures, then the attack will be executed to its ultimate goal of tripping generation and destabilizing the power system.
A physical-layer attack mitigation strategy is important. 
Even if attack detection measures fail, the SCC prevents the attack actions from tripping generation and alarms the system operator to the abnormal control.
The abnormal control can be logged for later inspection to reveal any system vulnerabilities or learn attacker strategies.
Next, the logged data can be used to develop tailored defense strategies based on understanding of attack strategies.

\subsection{Normal operation}

It is important to ensure that the SCC does not impede the operation of the power system. 
This includes ensuring that existing control continues to be able to maintain system security and stabilize system frequency after disturbances. 
First, we simulate a sudden large change in load in Fig. \ref{fig:hypersim_load_disturbance} to show that the SCC does not impede restoration of the frequency. 
The top plot in Fig. \ref{fig:hypersim_load_disturbance} shows the change in mechanical power output of the generator during the load change. 
The primary load-frequency control of the generator restores the frequency as seen in the bottom plot of Fig. \ref{fig:hypersim_load_disturbance}.
The SCC does not affect the reaction of the system to the load change.


Next, we simulate a 3-phase fault at bus B5 (see Fig. \ref{fig:mg_testbed}). The generator supplies the fault current to the fault. 
The phase A current along the transmission lines connecting bus B5 to bus B6 (to which the generator is connected) is shown in Fig. \ref{fig:hypersim_fault}. 
The over-current relay on the transmission line trips to isolate the fault. 
After the fault is cleared, existing control stabilizes the system back to normal operation. 
Throughout the process, the SCC does not affect existing control and protection, or the restoration of the power system. 


{
In assessing why the SCC does not impede restorative system operation, we note that the SCC controller has a slower reaction time compared to protection devices that are required to operate in  milliseconds. Hence, the SCC does not impede protection.
Moreover, following a sudden power change, there is no conflict between the SCC and existing stabilizing control, which maintains safety while stabilizing the system.
}

\section{Conclusion} \label{sec5:conclusion}

Cyberattackers can exploit the increased reliance of power systems on communication infrastructure and utilize knowledge of the control and protection of power systems to induce false relay operation with critical consequences including loss of generation, instability, cascading failure, and blackout. Power systems control relies on communicated signals to maintain system security; but in the case of an undetected cyberattack, adhering to the compromised signals can aid the cyberattacker's malicious goals. In this paper, we formulated a safety critical controller based on control barrier functions to mitigate false-data injection cyberattacks compromising load-frequency control. We demonstrated the controller's effectiveness in mitigating attacks targeting false relay operation of frequency and rate-of-change of frequency protection functions. We also showed that the controller does not impede normal restoration and stabilization procedures in power systems. {We validated the results using a high-fidelity testbed}.

\bibliography{references}
\end{document}